\documentclass[12pt]{iopart}
\usepackage{lineno}
\usepackage{amsmath, amssymb}
\usepackage{mathtools}

\usepackage[hidelinks, bookmarks=true]{hyperref}
\hypersetup{colorlinks=false,
            citebordercolor=1 1 1,
            linkbordercolor=1 1 1,
            pdfborderstyle={/S/N},
            pdfborder=0 0 0}
\usepackage{graphicx}
\usepackage{siunitx}
\usepackage{cite}
\newcommand{\addcauthor}[4]{#1$^{{#2},}$\footnotemark[{#3}]}
\newcommand{\addauthor}[2]{#1$^{#2}$}
\newcommand{\affiliation}[2]{\item[$^{#1}$] #2}

\begin{document}

% Line numbers for drafting
% \linenumbers

% Create title (square brackets for headers blank)
\title[]{Two-neutrino double electron capture of $^\mathbf{124}$Xe \newline in the first LUX-ZEPLIN exposure}

% Import pre-processed authorship list
\author{
{\large{The LUX-ZEPLIN (LZ) Collaboration}} \vspace{1.2mm} \\
% 1
\addauthor{J.~Aalbers}{1,2},
% 2
\addauthor{D.S.~Akerib}{1,2},
% 3
\addcauthor{A.K.~Al~Musalhi}{3}{1}{\href{mailto:aiham.almusalhi@ucl.ac.uk}{aiham.almusalhi@ucl.ac.uk}}, \addtocounter{footnote}{2}
% 4
\addauthor{F.~Alder}{3},
% 5
\addauthor{C.S.~Amarasinghe}{4,5},
% 6
\addauthor{A.~Ames}{1,2},
% 7
\addauthor{T.J.~Anderson}{1,2},
% 8
\addauthor{N.~Angelides}{6},
% 9
\addauthor{H.M.~Ara\'{u}jo}{6},
% 10
\addauthor{J.E.~Armstrong}{7},
% 11
\addauthor{M.~Arthurs}{1,2},
% 12
\addauthor{A.~Baker}{6},
% 13
\addauthor{S.~Balashov}{8},
% 14
\addauthor{J.~Bang}{9},
% 15
\addauthor{J.W.~Bargemann}{4},
% 16
\addauthor{E.E.~Barillier}{10,5},
% 17
\addauthor{K.~Beattie}{11},
% 18
\addauthor{A.~Bhatti}{7},
% 19
\addauthor{A.~Biekert}{11,12},
% 20
\addauthor{T.P.~Biesiadzinski}{1,2},
% 21
\addauthor{H.J.~Birch}{10,5},
% 22
\addauthor{E.~Bishop}{13},
% 23
\addauthor{G.M.~Blockinger}{14},
% 24
\addauthor{B.~Boxer}{15},
% 25
\addauthor{C.A.J.~Brew}{8},
% 26
\addauthor{P.~Br\'{a}s}{16},
% 27
\addauthor{S.~Burdin}{17},
% 28
\addauthor{M.~Buuck}{1,2},
% 29
\addauthor{M.C.~Carmona-Benitez}{18},
% 30
\addauthor{M.~Carter}{17},
% 31
\addauthor{A.~Chawla}{19},
% 32
\addauthor{H.~Chen}{11},
% 33
\addauthor{Y.T.~Chin}{18},
% 34
\addauthor{N.I.~Chott}{20},
% 35
\addauthor{M.V.~Converse}{21},
% 36
\addauthor{R.~Coronel}{1,2},
% 37
\addauthor{A.~Cottle}{3},
% 38
\addauthor{G.~Cox}{22},
% 39
\addauthor{D.~Curran}{22},
% 40
\addauthor{C.E.~Dahl}{24,23},
% 41
\addauthor{A.~David}{3},
% 42
\addauthor{J.~Delgaudio}{22},
% 43
\addauthor{S.~Dey}{25},
% 44
\addauthor{L.~de~Viveiros}{18},
% 45
\addauthor{L.~Di~Felice}{6},
% 46
\addauthor{C.~Ding}{9},
% 47
\addauthor{J.E.Y.~Dobson}{3},
% 48
\addauthor{E.~Druszkiewicz}{21},
% 49
\addauthor{S.~Dubey}{9},
% 50
\addauthor{S.R.~Eriksen}{26},
% 51
\addauthor{A.~Fan}{1,2},
% 52
\addauthor{N.M.~Fearon}{25},
% 53
\addauthor{N.~Fieldhouse}{25},
% 54
\addauthor{S.~Fiorucci}{11},
% 55
\addauthor{H.~Flaecher}{26},
% 56
\addauthor{E.D.~Fraser}{17},
% 57
\addauthor{T.M.A.~Fruth}{27},
% 58
\addauthor{R.J.~Gaitskell}{9},
% 59
\addauthor{A.~Geffre}{22},
% 60
\addauthor{J.~Genovesi}{20},
% 61
\addauthor{C.~Ghag}{3},
% 62
\addauthor{R.~Gibbons}{11,12},
% 63
\addauthor{S.~Gokhale}{28},
% 64
\addauthor{J.~Green}{25},
% 65
\addauthor{M.G.D.van~der~Grinten}{8},
% 66
\addauthor{J.J.~Haiston}{20},
% 67
\addauthor{C.R.~Hall}{7},
% 68
\addauthor{S.~Han}{1,2},
% 69
\addauthor{E.~Hartigan-O'Connor}{9},
% 70
\addauthor{S.J.~Haselschwardt}{11},
% 71
\addauthor{M.A.~Hernandez}{10,5},
% 72
\addauthor{S.A.~Hertel}{29},
% 73
\addauthor{G.~Heuermann}{5},
% 74
\addauthor{G.J.~Homenides}{30},
% 75
\addauthor{M.~Horn}{22},
% 76
\addauthor{D.Q.~Huang}{31},
% 77
\addauthor{D.~Hunt}{25},
% 78
\addauthor{E.~Jacquet}{6},
% 79
\addcauthor{R.S.~James}{3}{3}{Also at The University of Melbourne, School of Physics,
701 Melbourne, VIC 3010, Australia}, \addtocounter{footnote}{-1}
% 80
\addauthor{J.~Johnson}{15},
% 81
\addauthor{A.C.~Kaboth}{19},
% 82
\addauthor{A.C.~Kamaha}{31},
% 83
\addauthor{M.~Kannichankandy}{14},
% 84
\addauthor{D.~Khaitan}{21},
% 85
\addauthor{A.~Khazov}{8},
% 86
\addauthor{I.~Khurana}{3},
% 87
\addauthor{J.~Kim}{4},
% 88
\addauthor{Y.D.~Kim}{32},
% 89
\addauthor{J.~Kingston}{15},
% 90
\addauthor{R.~Kirk}{9},
% 91
\addauthor{D.~Kodroff}{18,11},
% 92
\addauthor{L.~Korley}{5},
% 93
\addauthor{E.V.~Korolkova}{33},
% 94
\addauthor{H.~Kraus}{25},
% 95
\addauthor{S.~Kravitz}{34},
% 96
\addauthor{L.~Kreczko}{26},
% 97
\addauthor{V.A.~Kudryavtsev}{33},
% 98
\addauthor{D.S.~Leonard}{32},
% 99
\addauthor{K.T.~Lesko}{11},
% 100
\addauthor{C.~Levy}{14},
% 101
\addauthor{J.~Lin}{11,12},
% 102
\addcauthor{A.~Lindote}{16}{2}{{\href{mailto:alex@coimbra.lip.pt}{alex@coimbra.lip.pt}}},
% 103
\addauthor{W.H.~Lippincott}{4},
% 104
\addauthor{M.I.~Lopes}{16},
% 105
\addauthor{W.~Lorenzon}{5},
% 106
\addauthor{C.~Lu}{9},
% 107
\addauthor{S.~Luitz}{1,2},
% 108
\addauthor{P.A.~Majewski}{8},
% 109
\addauthor{A.~Manalaysay}{11},
% 110
\addauthor{R.L.~Mannino}{35},
% 111
\addauthor{C.~Maupin}{22},
% 112
\addauthor{M.E.~McCarthy}{21},
% 113
\addauthor{G.~McDowell}{5},
% 114
\addauthor{D.N.~McKinsey}{11,12},
% 115
\addauthor{J.~McLaughlin}{23},
% 116
\addauthor{J.B.~McLaughlin}{3},
% 117
\addauthor{R.~McMonigle}{14},
% 118
\addauthor{E.~Mizrachi}{35,7},
% 119
\addauthor{A.~Monte}{4},
% 120
\addauthor{M.E.~Monzani}{1,2,36},
% 121
\addauthor{E.~Morrison}{20},
% 122
\addauthor{B.J.~Mount}{37},
% 123
\addauthor{M.~Murdy}{29},
% 124
\addauthor{A.St.J.~Murphy}{13},
% 125
\addauthor{A.~Naylor}{33},
% 126
\addauthor{H.N.~Nelson}{4},
% 127
\addauthor{F.~Neves}{16},
% 128
\addauthor{A.~Nguyen}{13},
% 129
\addauthor{C.L.~O'Brien}{34},
% 130
\addauthor{I.~Olcina}{11,12},
% 131
\addauthor{K.C.~Oliver-Mallory}{6},
% 132
\addauthor{J.~Orpwood}{33},
% 133
\addauthor{K.Y~Oyulmaz}{13},
% 134
\addauthor{K.J.~Palladino}{25},
% 135
\addauthor{J.~Palmer}{19},
% 136
\addauthor{N.J.~Pannifer}{26},
% 137
\addauthor{N.~Parveen}{14},
% 138
\addauthor{S.J.~Patton}{11},
% 139
\addauthor{B.~Penning}{10,5},
% 140
\addauthor{G.~Pereira}{16},
% 141
\addauthor{E.~Perry}{3},
% 142
\addauthor{T.~Pershing}{35},
% 143
\addauthor{A.~Piepke}{30},
% 144
\addauthor{Y.~Qie}{21},
% 145
\addauthor{J.~Reichenbacher}{20},
% 146
\addauthor{C.A.~Rhyne}{9},
% 147
\addauthor{Q.~Riffard}{11},
% 148
\addauthor{G.R.C.~Rischbieter}{10,5},
% 149
\addauthor{E.~Ritchey}{7},
% 150
\addauthor{H.S.~Riyat}{13},
% 151
\addauthor{R.~Rosero}{28},
% 152
\addauthor{T.~Rushton}{33},
% 153
\addauthor{D.~Rynders}{22},
% 154
\addauthor{D.~Santone}{19},
% 155
\addauthor{A.B.M.R.~Sazzad}{30},
% 156
\addauthor{R.W.~Schnee}{20},
% 157
\addauthor{G.~Sehr}{34},
% 158
\addauthor{B.~Shafer}{7},
% 159
\addauthor{S.~Shaw}{13},
% 160
\addauthor{T.~Shutt}{1,2},
% 161
\addauthor{J.J.~Silk}{7},
% 162
\addauthor{C.~Silva}{16},
% 163
\addauthor{G.~Sinev}{20},
% 164
\addauthor{J.~Siniscalco}{3},
% 165
\addauthor{R.~Smith}{11,12},
% 166
\addauthor{V.N.~Solovov}{16},
% 167
\addauthor{P.~Sorensen}{11},
% 168
\addauthor{J.~Soria}{11,12},
% 169
\addauthor{A.~Stevens}{3,6},
% 170
\addauthor{K.~Stifter}{24},
% 171
\addauthor{B.~Suerfu}{11,12},
% 172
\addauthor{T.J.~Sumner}{6},
% 173
\addauthor{M.~Szydagis}{14},
% 174
\addauthor{D.R.~Tiedt}{22},
% 175
\addauthor{M.~Timalsina}{11},
% 176
\addauthor{Z.~Tong}{6},
% 177
\addauthor{D.R.~Tovey}{33},
% 178
\addauthor{J.~Tranter}{33},
% 179
\addauthor{M.~Trask}{4},
% 180
\addauthor{M.~Tripathi}{15},
% 181
\addauthor{A.~Vacheret}{6},
% 182
\addauthor{A.C.~Vaitkus}{9},
% 183
\addauthor{O.~Valentino}{6},
% 184
\addauthor{V.~Velan}{11},
% 185
\addauthor{A.~Wang}{1,2},
% 186
\addauthor{J.J.~Wang}{30},
% 187
\addauthor{Y.~Wang}{11,12},
% 188
\addauthor{J.R.~Watson}{11,12},
% 189
\addauthor{L.~Weeldreyer}{30},
% 190
\addauthor{T.J.~Whitis}{4},
% 191
\addauthor{K.~Wild}{18},
% 192
\addauthor{M.~Williams}{5},
% 193
\addauthor{W.J.~Wisniewski}{1},
% 194
\addauthor{L.~Wolf}{19},
% 195
\addauthor{F.L.H.~Wolfs}{21},
% 196
\addauthor{S.~Woodford}{17},
% 197
\addauthor{D.~Woodward}{18,11},
% 198
\addauthor{C.J.~Wright}{26},
% 199
\addauthor{Q.~Xia}{11},
% 200
\addauthor{J.~Xu}{35},
% 201
\addauthor{Y.~Xu}{31},
% 202
\addauthor{M.~Yeh}{28},
% 203
\addauthor{D.~Yeum}{7},
% 204
\addauthor{W.~Zha}{18},
% 205
\addauthor{E.A.~Zweig}{31}
}
\address{
\begin{itemize}
\affiliation{1}{SLAC National Accelerator Laboratory, Menlo Park, CA 94025-7015, USA}
\affiliation{2}{Kavli Institute for Particle Astrophysics and Cosmology, Stanford University, Stanford, CA  94305-4085 USA}
\affiliation{3}{University College London (UCL), Department of Physics and Astronomy, London WC1E 6BT, UK}
\affiliation{4}{University of California, Santa Barbara, Department of Physics, Santa Barbara, CA 93106-9530, USA}
\affiliation{5}{University of Michigan, Randall Laboratory of Physics, Ann Arbor, MI 48109-1040, USA}
\affiliation{6}{Imperial College London, Physics Department, Blackett Laboratory, London SW7 2AZ, UK}
\affiliation{7}{University of Maryland, Department of Physics, College Park, MD 20742-4111, USA}
\affiliation{8}{STFC Rutherford Appleton Laboratory (RAL), Didcot, OX11 0QX, UK}
\affiliation{9}{Brown University, Department of Physics, Providence, RI 02912-9037, USA}
\affiliation{10}{University of Zurich, Department of Physics, 8057 Zurich, Switzerland}
\affiliation{11}{Lawrence Berkeley National Laboratory (LBNL), Berkeley, CA 94720-8099, USA}
\affiliation{12}{University of California, Berkeley, Department of Physics, Berkeley, CA 94720-7300, USA}
\affiliation{13}{University of Edinburgh, SUPA, School of Physics and Astronomy, Edinburgh EH9 3FD, UK}
\affiliation{14}{University at Albany (SUNY), Department of Physics, Albany, NY 12222-0100, USA}
\affiliation{15}{University of California, Davis, Department of Physics, Davis, CA 95616-5270, USA}
\affiliation{16}{{Laborat\'orio de Instrumenta\c c\~ao e F\'isica Experimental de Part\'iculas (LIP)}, University of Coimbra, P-3004 516 Coimbra, Portugal}
\affiliation{17}{University of Liverpool, Department of Physics, Liverpool L69 7ZE, UK}
\affiliation{18}{Pennsylvania State University, Department of Physics, University Park, PA 16802-6300, USA}
\affiliation{19}{Royal Holloway, University of London, Department of Physics, Egham, TW20 0EX, UK}
\affiliation{20}{South Dakota School of Mines and Technology, Rapid City, SD 57701-3901, USA}
\affiliation{21}{University of Rochester, Department of Physics and Astronomy, Rochester, NY 14627-0171, USA}
\affiliation{22}{South Dakota Science and Technology Authority (SDSTA), Sanford Underground Research Facility, Lead, SD 57754-1700, USA}
\affiliation{23}{Northwestern University, Department of Physics \& Astronomy, Evanston, IL 60208-3112, USA}
\affiliation{24}{Fermi National Accelerator Laboratory (FNAL), Batavia, IL 60510-5011, USA}
\affiliation{25}{University of Oxford, Department of Physics, Oxford OX1 3RH, UK}
\affiliation{26}{University of Bristol, H.H. Wills Physics Laboratory, Bristol, BS8 1TL, UK}
\affiliation{27}{The University of Sydney, School of Physics, Physics Road, Camperdown, Sydney, NSW 2006, Australia}
\affiliation{28}{Brookhaven National Laboratory (BNL), Upton, NY 11973-5000, USA}
\affiliation{29}{University of Massachusetts, Department of Physics, Amherst, MA 01003-9337, USA}
\affiliation{30}{University of Alabama, Department of Physics \& Astronomy, Tuscaloosa, AL 34587-0324, USA}
\affiliation{31}{University of California, Los Angeles, Department of Physics \& Astronomy, Los Angeles, CA 90095-1547, USA}
\affiliation{32}{IBS Center for Underground Physics (CUP), Yuseong-gu, Daejeon, Korea}
\affiliation{33}{University of Sheffield, Department of Physics and Astronomy, Sheffield S3 7RH, UK}
\affiliation{34}{University of Texas at Austin, Department of Physics, Austin, TX 78712-1192, USA}
\affiliation{35}{Lawrence Livermore National Laboratory (LLNL), Livermore, CA 94550-9698, USA}
\affiliation{36}{Vatican Observatory, Castel Gandolfo, V-00120, Vatican City State}
\affiliation{37}{Black Hills State University, School of Natural Sciences, Spearfish, SD 57799-0002, USA}
\end{itemize}
}

% Manually treat footnotes for corresponding authors
\addtocounter{footnote}{-1}
\footnotetext[1]{\href{mailto:aiham.almusalhi@ucl.ac.uk}{aiham.almusalhi@ucl.ac.uk}}
\addtocounter{footnote}{1}
\footnotetext[2]{\href{mailto:alex@coimbra.lip.pt}{alex@coimbra.lip.pt}}
\addtocounter{footnote}{1}
\footnotetext[3]{Also at The University of Melbourne, School of Physics, Melbourne, VIC 3010, Australia}

% Generate the abstract
\begin{abstract}
The broad physics reach of the LUX-ZEPLIN (LZ) experiment covers rare phenomena beyond the direct detection of dark matter. We report precise measurements of the extremely rare decay of $^{124}$Xe through the process of two-neutrino double electron capture (2$\nu$2EC), utilizing a $\qty{1.39}{kg \times yr}$ isotopic exposure from the first LZ science run. A half-life of $T_{1/2}^{\hspace{0.3mm}2\nu2\mathrm{EC}} = (1.09 \pm 0.14_{\text{stat}} \pm 0.05_{\text{sys}}) \times \qty{e22}{yr}$ is observed with a statistical significance of \qty{8.3}{\sigma}, in agreement with literature. First empirical measurements of the KK capture fraction relative to other K-shell modes were conducted, and demonstrate consistency with respect to recent signal models at the \qty{1.4}{\sigma} level.
\end{abstract}

%\maketitle

% -------------------------------------------------------------

\section{Introduction}

Alongside a world-leading sensitivity to keV-scale scatters from weakly interacting massive particles (WIMPs), dual-phase xenon time projection chambers (TPCs) are also capable of searching for rare phenomena beyond dark matter interactions \cite{benetti, cline}. Most notably, this encompasses experimental probes for the fundamental nature of the neutrino mass such as neutrinoless double beta decay ($0\nu\beta\beta$), the observation of which may indicate that neutrinos are Majorana particles \cite{majorana, delloro_rev, dolinski_rev, agostini_rev}. Neutrinoless double electron capture ($0\nu$2EC) is an analogous process that would be associated with the proton-rich side of the binding energy parabola for even-even isobars \cite{bernabeu, sujkowski_0v2EC, vsimkovic}; $0\nu$2EC decays have yet to be observed, with extremely long expected half-lives in excess of \qty{e29}{yr} \cite{Wittweg}. Models and predictions for this process may be developed through measurements of the adjacent process of two-neutrino double electron capture (2$\nu$2EC) \cite{suhonen}, which involves the absorption of two atomic electrons and the subsequent conversion of a pair of protons into neutrons, along with the simultaneous emission of two neutrinos. Measured half-lives for this decay inform the underlying nuclear matrix element $M^{(2\nu)}$ through the relation
\begin{equation}
    \left(T_{1/2}^{\hspace{0.3mm}2\nu2\mathrm{EC}}\right)^{-1} = G_{2\nu}^{2\mathrm{EC}} g^{4}_{A} 
    \left|m_e c^2 M^{(2\nu)}\right|^{2},
\end{equation}

\noindent where $g_A$ denotes the weak axial-vector coupling strength, $m_e$ is the electron mass, and the phase space factor $G_{2\nu}^{2\mathrm{EC}}$ is proportional to the reaction Q-value taken to the fifth power \cite{HL_2v2EC}. Thus far, 2$\nu$2EC has only been observed for $^{130}$Ba \cite{Ba130a, Ba130b}, $^{78}$Kr \cite{baksan1, baksan2}, and $^{124}$Xe \cite{XENON1T_Xe124_KK, XENON1T_DEC}.

In this work, the extensive physics program of the LUX-ZEPLIN (LZ) experiment is highlighted with a precise measurement of the $^{124}$Xe decay half-life through 2$\nu$2EC. This result is extended with the first measurement of the relative capture fractions between atomic shell combinations. A brief description of the LZ experiment and its first exposure is provided in section \ref{sec:LZ}. This is followed by a detailed overview of the analysis given in section \ref{sec:analysis}, which covers the employed signal model, construction of the background model with an emphasis on the dominant $^{125}$I background, and an outline of the fit procedure. Finally, results of the analysis, including the measured relative capture fractions, are discussed in section \ref{sec:results}.

% -------------------------------------------------------------

\section{The LUX-ZEPLIN experiment}
\label{sec:LZ}

The LZ experiment \cite{LZ_NIM, LZ_TDR} is situated \qty{1480}{m} underground within the Davis Cavern at the Sanford Underground Research Facility in Lead, South Dakota. With a \qty{1100}{m} ($\qty{4300}{\mathrm{m.w.e}}$) rock overburden, the flux of cosmic muons in the cavern is attenuated by a factor of \qty{3e6}{} with respect to that at the surface \cite{sanford, muons}. The detector is housed within a tank holding \qty{238}{t} of ultra-pure water for additional shielding from ambient radiation. Combined with an extensive radioassay campaign and strict cleanliness protocols in assembly, this establishes an ultra-low background environment~\cite{LZ_assays}.

At the core of the LZ detector is a dual-phase xenon TPC with an active volume containing \qty{7}{t} of liquid xenon (LXe). Energy depositions in the LXe generate vacuum ultraviolet scintillation photons (S1), as well as ionization electrons. The electrons drift to the liquid surface under a vertical electric field and are subsequently extracted into the gas phase by a stronger field, where they produce a secondary scintillation signal (S2). Light is detected by two arrays of 3-inch photomultiplier tubes (PMTs), with 253 at the top and 241 at the bottom. To enhance background rejection, the detector features two anti-coincidence veto systems: an optically isolated LXe skin surrounding the TPC, which is instrumented with 93 1-inch and 38 2-inch PMTs; and a near-hermetic outer detector (OD) comprised of acrylic tanks filled with gadolinium-loaded liquid scintillator, viewed by 120 8-inch PMTs in the water tank. These detectors enable the rejection of events with coincident gamma rays or neutrons.

The detector response is calibrated with a variety of dedicated sources \cite{LZ_calibs}. In particular, dispersed mono-energetic sources such as $^{83\mathrm{m}}$Kr and $^{131\mathrm{m}}$Xe are used to correct the S1 and S2 observables for position-dependent effects; the corrected versions are denoted as S1c and S2c, and are measured in units of photons detected (phd) \cite{LZ_SR1}. These quantities are combined to reconstruct the energy of an event according to
\begin{equation}
    E = W \left(\frac{\mathrm{S1c}}{g_1} + \frac{\mathrm{S2c}}{g_2}\right),
\end{equation}

\noindent where $g_1 = \qty{0.114(2)}{phd/photon}$ and $g_2 = \qty{47.1(11)}{phd/electron}$ respectively describe the photon detection efficiency and the effective charge gain, and $W = \qty{13.5}{eV}$ is the assumed work function \cite{W_val}.

Of the calibration sources deployed during the first exposure of LZ, a \qty{4.5}{d} calibration campaign partway through the run is of particular relevance to this work. This utilized neutrons generated by deuterium-deuterium (DD) fusion to characterize the nuclear recoil response of the detector \cite{DD_brown, DD_LUX}. The deployment of neutron calibrations introduces additional transient backgrounds from activation of xenon isotopes. Treatment of these sources for the background model is detailed in section \ref{sec:bg}.

% -------------------------------------------------------------

\section{Analysis overview}
\label{sec:analysis}

The data analyzed here are from Science Run 1 (SR1), acquired between December 23rd, 2021, and May 11th, 2022. For the first LZ WIMP search (WS) result \cite{LZ_SR1}, a suite of data quality cuts were developed to target accidental coincidence backgrounds, which originate from false pairings of uncorrelated S1s and S2s, as well as other unique detector pathologies such as photon pileup and delayed electron emission. A number of these cuts involved the exclusion of periods with elevated detector activity, and hence posed a substantial impact on the accrued live time. For the purpose of this analysis, the contribution of such effects is not significant as the focus is at a higher energy regime, and hence only a minimal set of cuts need be applied: a fiducial volume (FV) cut to reject external backgrounds near the TPC boundaries; pulse area thresholds of $\left(\mathrm{S1c} > \qty{100}{phd}\right)$ and $\left(\mathrm{S2} > \qty{600}{phd}\right)$ to mitigate for tails of accidental backgrounds; and vetoes from the skin and OD to tackle events with multiple interaction vertices. This results in an overall exposure of \qty{96.4(10)}{} live days. The same \qty{5.5(2)}{t} FV definition was maintained as the SR1 WS analysis. Constraints on various background rates in the FV were informed by dedicated measurements from several studies~\cite{LZ_SR1BG}.

\subsection{Signal model}

For $^{124}$Xe, 2$\nu$2EC proceeds as
\begin{equation}
    {}^{124}\mathrm{Xe} + 2e^{-} \longrightarrow {}^{124}\mathrm{Te} + 2\nu_{e} + (\text{X-rays \& Auger electrons}),
\end{equation}

\noindent with a reaction Q-value of $\qty{2856.73(12)}{keV}$ \cite{Xe124_Qval}. Most of this energy is carried away by the neutrinos, whilst the nuclear recoil energy is only $\mathcal{O}(10)\: \text{eV}$ \cite{LUX_Xe124}, and hence the only detectable signal consists entirely of deposits from the X-rays and Auger cascade in the subsequent atomic de-excitation of the $^{124}$Te daughter. The energy of the experimental signature is determined by the combination of atomic shells from which electrons were captured. Here we assume the same relative shell capture fractions and energies as the XENON collaboration \cite{XENON1T_DEC}; these are summarized in table \ref{table:fractions}. All capture combinations are expected to manifest as single-scatter (SS) interactions due to the sub-millimeter reach in LXe at the energy scale of the emitted X-rays and Auger electrons.

The half-life $T_{1/2}^{\hspace{0.3mm}2\nu2\mathrm{EC}}$ is related to the activity $A_{2\nu2\mathrm{EC}}$ by
\begin{equation}
    \label{eq:activity}
    T_{1/2}^{\hspace{0.3mm}2\nu2\mathrm{EC}} = \frac{\eta N_{A}}{A_{2\nu2\mathrm{EC}} M_{A}} \ln{2},
\end{equation}

\noindent where $N_A$ is Avogadro's constant, $M_A = \qty{131.293(6)}{g \per mol}$ is the molar mass of xenon \cite{Xe_IUPAC}, and $\eta$ is the isotopic abundance of $^{124}$Xe. It is assumed that the isotopic composition of xenon in LZ is consistent with that of atmospheric xenon. At the natural $^{124}$Xe abundance $\eta = \qty{0.0952(3)}{\%}$ \cite{Xe_abundance}, the isotopic exposure is $\qty{1.39}{kg \times yr}$. Out of ${\sim}400$ decays over the course of this exposure (for $T_{1/2}^{\hspace{0.3mm}2\nu2\mathrm{EC}} = \qty{1.1e22}{yr}$ \cite{XENON1T_DEC}), $<10$ events are expected to occur from double L-shell (LL) and higher-shell electron captures. Therefore, a 20--100 keV interval was set as the region of interest (ROI) for this search; the lower bound reflects the lack of sensitivity to these modes, whereas the upper bound encapsulates the energies of flat backgrounds extending beyond the signal peaks.

\begin{table}[!htb]
\caption{\label{table:fractions} Energies and relative capture fractions for different shell combinations in the $^{124}$Xe 2$\nu$2EC signal model, as adopted in \cite{XENON1T_DEC}. Decay modes are labeled by the combination of shells from which electrons were captured.}
    \begin{indented}
        \item[]\begin{tabular}{@{}lcc}
        \br
        Decay mode & Capture fraction [\%] & Energy [keV] \\
        \mr
        KK & 72.4 & 64.3 \\
        KL & 20.0 & 36.7–37.3 \\
        KM & 4.3 & 32.9–33.3 \\
        KN & 1.0 & 32.3–32.4 \\
        LL & 1.4 & 8.8–10.0 \\
        Other & 0.9 & $<10$ \\
        \br
        \end{tabular}
    \end{indented}
\end{table}
\normalsize

% -------------------------------------------------------------

\subsection{Background model}
\label{sec:bg}

The majority of backgrounds within the 20--\qty{100}{keV} 2$\nu$2EC ROI, listed in table \ref{table:backgrounds}, exhibit flat and continuous energy spectra, enabling straightforward modeling. Furthermore, their rates in SR1 have been well-constrained by a variety of sideband studies, as reported in reference \cite{LZ_SR1BG}. A subset of these backgrounds is comprised of unstable xenon isotopes, which were produced via cosmogenic activation during transport, as well as through neutron activation as a result of deploying neutron sources to calibrate the detector response to nuclear recoils. Aside from $^{133}$Xe, which features as a beta spectrum with a shoulder starting at the \qty{81.0}{keV} gamma-ray line \cite{A133}, these can deposit energies within close proximity to the secondary (KL + KM + KN) $^{124}$Xe signal peak, denoted here as KX. Isotopes with decays involving a single electron capture, namely $^{125}$Xe and $^{127}$Xe, involve cascades with a total deposited energy of \qty{33.2}{keV} for K-shell captures \cite{A127}. However, these decays are accompanied by a gamma ray, which often gives rise to a multiple-scatter (MS) topology or is tagged by the veto detectors, with an efficiency of \qty{78.0(50)}{\%}, and hence they are suppressed by the SS and veto cuts. Similarly, $^{129\mathrm{m}}$Xe emits a pair of coincident \qty{196.6}{keV} and \qty{39.6}{keV} internal conversion electrons in its transitions to the ground state \cite{A129}, with a very low probability of detecting the lower energy gamma ray in isolation.

\begin{table}
\caption{\label{table:backgrounds} Summary of backgrounds to the $^{124}$Xe $2\nu2$EC measurement. Reported rate estimates are for SS events within the ROI, after the application of area thresholds, vetoes, and the FV cut. Solar neutrino and detector material backgrounds consist of a sum over all relevant sources. K-shell EC and single gamma-ray rates are for cases where the skin veto fails to tag an outgoing associated gamma ray; $^{125}$Xe is excluded as its average rate is effectively zero in this context. Half-lives are quoted from nuclear data sheets \cite{A125, A127, A129, A133}, and otherwise omitted in cases where they exceed the lifetime of the experiment, or for sources that are continually produced.}
    \begin{indented}
        \item[]\begin{tabular}{@{}lccc}
        \br
        Source & ROI signature & Half-life & Average rate [mHz] \\
        \mr
        \vspace{-3.8mm} \rule{0pt}{1pt} & & & \\
        $^{125}$I & Multiple EC peaks & \qty{59.4}{d} & \qty{1.75(38)e-1}{} \\
        $^{127}$Xe & K-shell EC (\qty{33.2}{keV}) & \qty{36.3}{d} & \qty{3.98(25)e-3}{} \\
        $^{129\mathrm{m}}$Xe & $\gamma$ (\qty{39.6}{keV}) & \qty{8.9}{d} & \qty{7.31(52)e-4}{} \\
        $^{133}$Xe & $\beta + \gamma$ (\qty{81.0}{keV}) & \qty{5.2}{d} & \qty{7.36(75)e-1}{} \\
        $^{136}$Xe $2\nu\beta\beta$ & continuous & - & \qty{1.34(32)e-1}{} \\
        $^{212}$Pb & continuous & - & \qty{5.70(3)e-3}{} \\
        $^{214}$Pb & continuous & - & \qty{2.16(5)e-1}{} \\
        $^{85}$Kr & continuous & - & \qty{3.83(92)e-2}{} \\
        Solar neutrinos & continuous & - & \qty{2.71(7)e-2}{} \\
        Materials & continuous & - & \qty{8.25(3)e-4}{}\\
        \br
        \end{tabular}
    \end{indented}
\end{table}
\normalsize

An especially problematic background is $^{125}$I, which is produced by the neutron activation of $^{124}$Xe according to the following steps
\begin{align}
    {}^{124}\text{Xe} + n &\longrightarrow {}^{125}\text{Xe} + \gamma, \\
    {}^{125}\text{Xe} + e^{-} &\mathrel{\overset{\text{\scriptsize{EC}}}{\underset{\text{\scriptsize{\qty{16.9}{h}}}}{\longrightarrow}}} {}^{125}\text{I} + \nu_e + \gamma + (\text{X-rays \& Auger electrons}),
\end{align}

\noindent and decays to an excited state of $^{125}$Te via electron capture as
\begin{equation}
    {}^{125}\text{I} + e^{-} \mathrel{\overset{\text{\scriptsize{EC}}}{\underset{\text{\scriptsize{\qty{59.4}{d}}}}{\longrightarrow}}} {}^{125}\text{Te} + \nu_e + \gamma + (\text{X-rays \& Auger electrons}).
\end{equation}

The combination of the subsequent de-excitation cascades and the \qty{35.5}{keV} nuclear transition of the $^{125}$Te daughter generates peaks that significantly overlap with the $^{124}$Xe signals. For instance, the K-shell capture mode of $^{125}$I has an associated total energy of \qty{67.3}{keV}; with a measured energy resolution of \qty{4.5(4)}{\%}, this peak is \qty{1}{\sigma} from the \qty{64.3}{keV} KK peak $^{124}$Xe. This is similarly the case for the other $^{125}$I decay modes, as shown in table \ref{table:I125_BRs}.

\begin{table} [!htb]
\caption{\label{table:I125_BRs} Energies and capture fractions for $^{125}$I background peaks. The capture fractions are quoted from \cite{atomic_datasheet}, and the binding energies are taken from \cite{Xray_trans1, Xray_trans2} with uncertainties excluded as they are of $\mathcal{O}(1)\: \text{eV}$.}
    \begin{indented}
        \item[]\begin{tabular}{@{}lll}
        \br
        Decay mode & Capture fraction [\%] & Energy [keV] \\
        \mr
        $\gamma$ + K & $80.11 \pm 0.17$ & 67.3 (35.5 + 31.8) \\
        $\gamma$ + L & $15.61 \pm 0.13$ & 40.4 (35.5 + 4.9)\\
        $\gamma$ + M & $3.49 \pm 0.07$ & 36.5 (35.5 + 1.0)\\
        \br
        \end{tabular}
    \end{indented}
\end{table}
\normalsize

In SR1, $^{125}$I was predominantly produced during the DD calibration campaign. Whilst it has a natural half-life of \qty{59.4}{d} \cite{A125}, it is efficiently removed by the LZ online purification system, which employs a hot zirconium getter \cite{LZ_NIM}. To quantify the removal rate during SR1, the \qty{67.3}{keV} peak associated with K-shell ECs in $^{125}$I was selected in data immediately following the DD calibration using the \qty{2}{\sigma} contour of a two-dimensional Gaussian fit of the corresponding population in $(\mathrm{S1c}, \mathrm{S2c})$ space. Events passing this selection along with the specified minimal set of cuts were divided into one-day bins according to their registered trigger timestamps, with the rate calculated in each bin adjusted for the experimental dead time.

\begin{figure}[!htb]
    \centering
    \includegraphics[width=\linewidth]{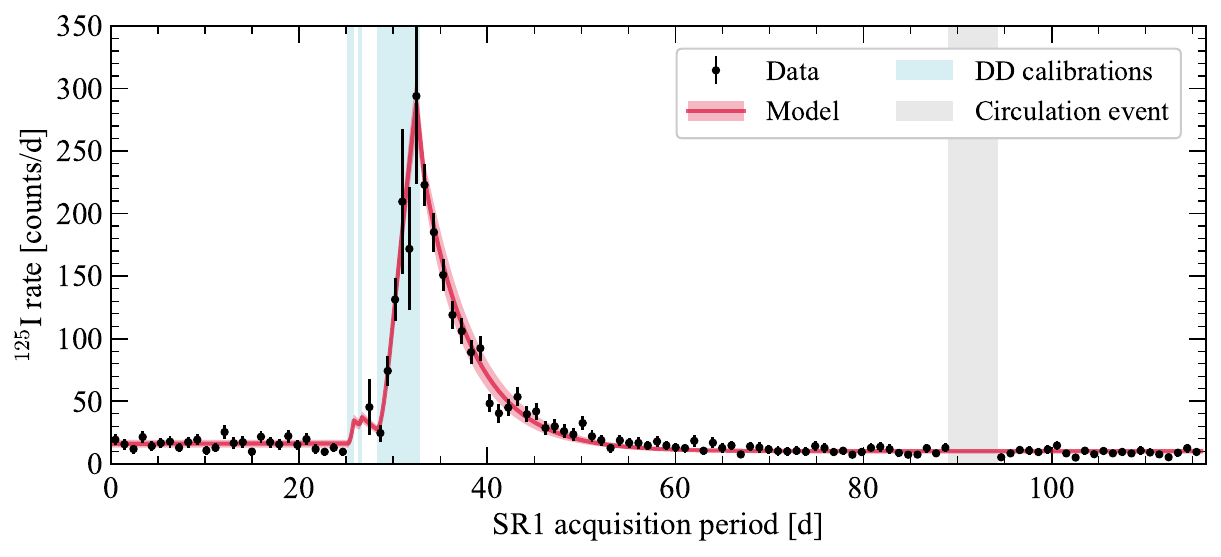}
    \caption{Rate of events in the FV over SR1 for a selection dominated by $^{125}$I. Along with skin and OD vetoes, a \qty{2}{\sigma} elliptical contour was applied from a two-dimensional Gaussian fit to the $^{125}$I K-shell EC population in (S1c, S2c) space. The overlaid profile represents a combination of the neutron activation model and a fit to the post-DD decay trend. The additional intermission window beyond the DD calibration period was caused by a circulation event, over which the detector purity temporarily decreased.}
    \label{fig:I125removal}
\end{figure}

The observed time profile of $^{125}$I rates over the course of SR1 is visualized in figure~\ref{fig:I125removal}, with a pronounced growth due to neutron activation following the DD calibrations. To model the behavior of its rate in the post-DD period, the number of $^{125}$I nuclei at a given time, $N_{^{125}\mathrm{I}}(t)$, was described by a differential equation
\begin{equation}
    \frac{dN_{^{125}\mathrm{I}}(t)}{dt} = \lambda_{^{125}\mathrm{Xe}} N_{^{125}\mathrm{Xe}}(t) - \left(\lambda_{^{125}\mathrm{I}} + \lambda_{g}\right) N_{^{125}\mathrm{I}}(t),
\end{equation}

\noindent in which the positive term represents the production of $^{125}$I from the decay of $^{125}$Xe, and the effective decay constant $\lambda_{\mathrm{eff}} = \left(\lambda_{^{125}\mathrm{I}} + \lambda_{g}\right)$ consists of a getter removal component $\lambda_g$ and a natural decay component $\lambda_{^{125}\mathrm{I}}$. A fit to the post-DD exponential decay in figure \ref{fig:I125removal} yields a getter removal half-life of $t_{1/2}^{g} = \qty{3.8(2)}{d}$, which translates into an effective $^{125}$I half-life of $t_{1/2}^{\mathrm{eff}} = \qty{3.6(2)}{d}$ with the natural decay component included. For reference, the values reported for $t_{1/2}^{g}$ by LUX and XENON1T are \qty{3.7(3)}{d} and \qty{4.6(1.6)}{d}, respectively \cite{LUX_Xe124, XENON1T_DEC}. In each case, the corresponding getter removal time constant was consistent with the interval for xenon to flow through one cycle of the circulation system.

In addition, the expected amount of $^{125}$I was also constrained by an activation model. The activation rate was characterized using DD calibration data by profiling the rates of both $^{125}$I and $^{125}$Xe within the vicinity of the neutron conduit at the top of the TPC. This subsequently informed the predicted $^{125}$I rate through the DD period given the measured getter removal time constant. The projected rate within the post-DD period is in very close agreement with the direct fit of the decay trend, and hence the latter is adopted for the combined model curve overlaid in figure \ref{fig:I125removal}.

In general, the baseline rate was expected to consist of continuous backgrounds, some contamination from the $2\nu$KK mode of $^{124}$Xe, and a flat rate of $^{125}$I from thermal neutron activation. However, a shift was observed between the pre-DD and post-DD baseline rates, from \qty{16.1(32)}{counts\per d} down to \qty{10.0(0.5)}{counts\per d}. This was attributed to a residual $^{125}$I population in the pre-DD window from activation during longer neutron calibrations performed a month prior to the start of SR1. With different circulation settings used in the physics commissioning period leading up to SR1, such as a higher flow rate through the getter, the $^{125}$I removal efficiency was slightly lower and hence extended the pre-DD decay curve. Nevertheless, the low pre-DD $^{125}$I rate is sufficiently well-approximated as flat over time.

% -------------------------------------------------------------

\subsection{Fit procedure}

Since the signals and backgrounds both produce electron recoils (ERs), reconstructed energy was adopted as the main observable. Furthermore, rather than excluding a post-DD window with elevated background rates, the full exposure can be preserved by leveraging the temporal variation of the primary $^{125}$I background, which motivates the use of calendar time as an additional observable. Given the substantial overlap in energy between $^{125}$I and the signal, incorporating this temporal information leads to a boost in sensitivity.

Two-dimensional fits in energy and time were performed according to a binned extended maximum likelihood approach, using a likelihood function of the form
\begin{equation}
    \ln{\mathcal{L}(\boldsymbol{d}|\boldsymbol{\theta},\boldsymbol{\mu},\boldsymbol{\sigma})} = \sum_{i}^{N_{E}} \sum_{j}^{N_{t}} d_{ij} \ln f_{ij}(\boldsymbol{\theta})
    - \nu(\boldsymbol{\theta})
    -\sum_{k}^{N_c} \frac{1}{2}\left(\frac{\theta_k - \mu_k}{\sigma_k}\right)^{2},
\end{equation}

\noindent where observed counts $d_{ij}$ and expected counts $f_{ij}(\boldsymbol{\theta})$ are divided into $N_E$ \qty{1}{keV} energy bins and $N_t$ \qty{1}{d} time bins. For each of the $N_c$ signal plus background components, there is an associated energy PDF $p_k$ and time PDF $t_k$, both of which are scaled by the corresponding rate parameter $\theta_k$ such that $f_{ij}(\boldsymbol{\theta}) = \sum^{N_c}_{k} \theta_{k} p_{ik} t_{jk}$. Energy spectra produced by the LZ parametric simulations chain were normalized to produce $p_k$ \cite{LZ_sims}, whereas $t_k$ represents the live time distribution over calendar time scaled by rate profile of the $k$th component, which is either constant or an exponential decay. The extended Poisson term $\nu(\boldsymbol{\theta}) = \sum_{i}^{N_{E}} \sum_{j}^{N_t} f_{ij}(\boldsymbol{\theta})$ serves to include information from the total sample size.

To simplify the fit procedure, the SR1 dataset was split into pre-DD and post-DD subsets consisting of 23.1 and 73.3 live days each, and evaluated with a simultaneous likelihood fit. The model contains a total of 10 parameters: 6 are shared between the two likelihoods, and a pair of unique pre-DD and post-DD rates is associated with each of $^{125}$I and $^{133}$Xe as they are impacted by the DD calibrations. Rates for $^{212}$Pb and detector materials were fixed as they were comparatively subdominant.

The model parameters $\boldsymbol{\theta}$ are regularized by a Gaussian constraint term, which penalizes significant deviations from the expected rates $\boldsymbol{\mu}$ as a fraction of their uncertainties $\boldsymbol{\sigma}$. Whereas the signal term was left unconstrained, initial estimates for background rates and their constraints were derived from the results of sideband analyses \cite{LZ_SR1BG}. In particular, tight constraints on $^{125}$I were enabled by the activation model outlined in section \ref{sec:bg}. Goodness of fit is quantified in terms of the metric $\chi^{2}_{\lambda} = 2 \sum_{i}^{N_{E}} \sum_{j}^{N_{t}} f_{ij}(\boldsymbol{\theta}) \hspace{0.85pt} - \hspace{0.85pt} d_{ij} \hspace{0.85pt} + \hspace{0.85pt} d_{ij} \ln(d_{ij} / f_{ij}(\boldsymbol{\theta}))$, which is constructed by means of a Poisson likelihood ratio test \cite{chisq}.

% -------------------------------------------------------------

\section{Results and discussion}
\label{sec:results}

\begin{figure}[p]
    \centering
    % \vspace{5mm}
    \includegraphics[width=\linewidth]{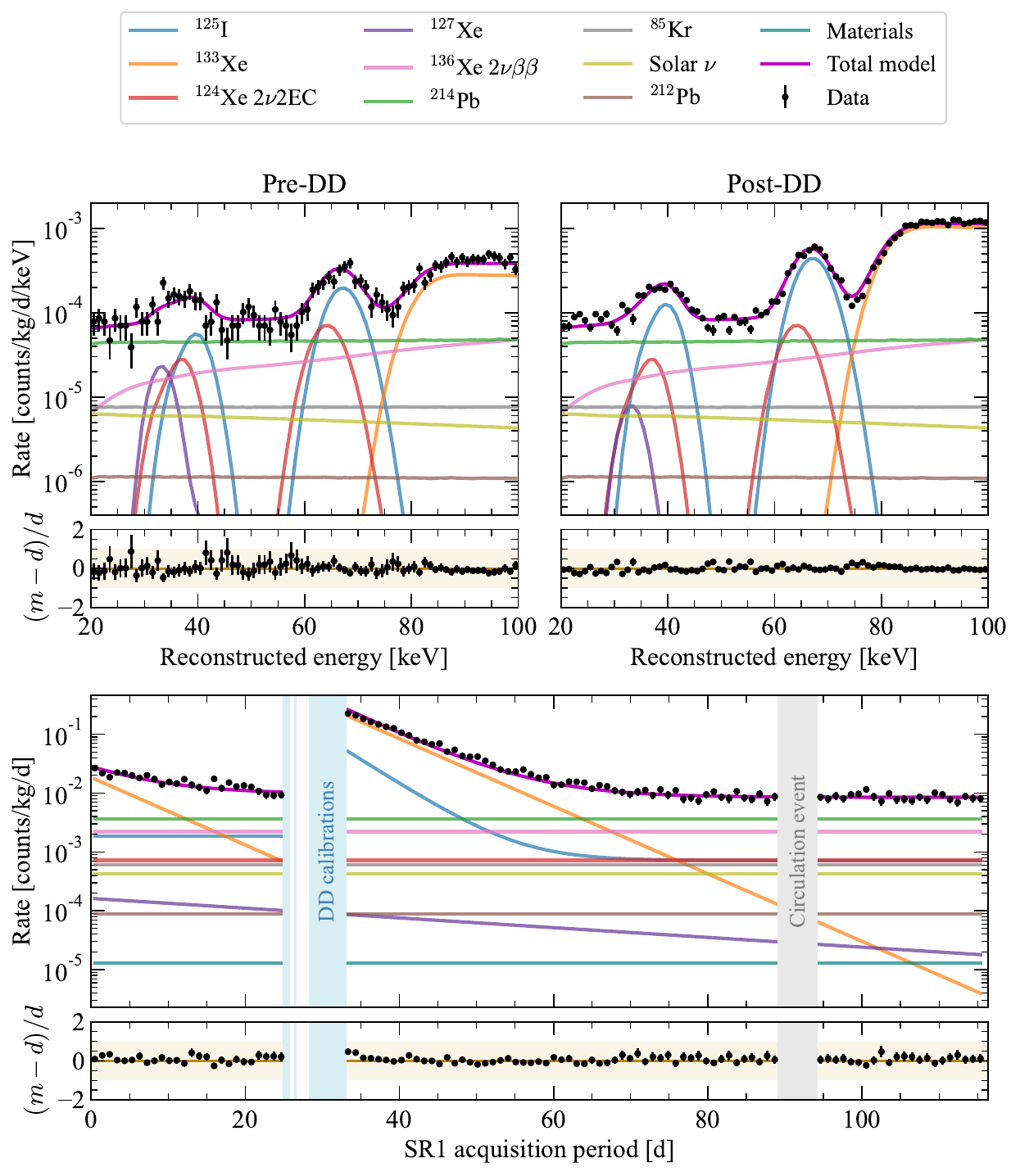}
    \caption{Simultaneous fit of pre-DD and post-DD subsets of SR1 projected into energy and time. Both $^{212}$Pb and the detector material components were fixed as they are subdominant. The top panel is displayed with rates averaged over the pre-DD and post-DD time windows, whereas rates in the bottom panel were integrated over the entire ROI.  Although $^{133}$Xe appears to dominate in the time projection plot, it is well-separated from the signal peaks in energy. The residual plots show fractional differences between the best-fit model $\boldsymbol{m}$ and the data $\boldsymbol{d}$ to illustrate the fit quality, with the most significant differences occurring in the low-statistics regions between features. The overall $\chi^{2}_{\lambda}/N_{\text{dof}} = 4451.22/4462 = 1.00$ is indicative of a high quality fit. The energy-only goodness of fit, calculated by marginalizing over time, is $\chi^{2}_{\lambda}/N_{\text{dof}} = 197.34/150 = 1.32$.}
    \label{fig:fit}
\end{figure}

The outcome of the fit is presented in figure \ref{fig:fit} as separate projections in energy and time, with a best-fit activity of $A_{2\nu 2\text{EC}} = \qty{0.28(4)}{counts\per\kilogram\per yr}$. Evaluation of a likelihood ratio test statistic with respect to the the background-only case yields a significance of \qty{8.3}{\sigma} for the presence of a 2$\nu$2EC signal. The half-life inferred using equation \ref{eq:activity} is
\begin{equation}
    T_{1/2}^{\hspace{0.3mm}2\nu2\mathrm{EC}} = (1.09 \pm 0.14_{\text{stat}} \pm 0.05_{\text{sys}}) \times \qty{e22}{yr},
\end{equation}

\noindent which is in excellent agreement with the previously reported measurement by XENON1T at $T_{1/2}^{\hspace{0.3mm}2\nu2\mathrm{EC}} = (1.1 \pm 0.2_{\text{stat}} \pm 0.1_{\text{sys}}) \times \qty{e22}{yr}$ \cite{XENON1T_DEC}. A breakdown of the systematic uncertainties is given in table \ref{table:systematics}. The dominant contribution is formed by the uncertainty on the FV mass. Secondary sources of uncertainty stem from the energy reconstruction parameters $g_{1}$ and $g_{2}$ as calibrated for the WS, as well as the live time estimate.

% "although 133Xe decays also exhibit a similar time dependence, 125I decays dominate the KK energy region which allows the rate to be accurately determined" either in the caption or in the text.

\begin{table} [!htb]
\caption{\label{table:systematics} Summary of systematic uncertainties on the $^{124}$Xe $2\nu2$EC half-life, with the total calculated by summing the individual contributions in quadrature.}
    \begin{indented}
        \item[]\begin{tabular}{@{}lcc}
        \br
        Contribution & Uncertainty [\qty{e22}{yr}] & Relative uncertainty [\%] \\
        \mr
        Fiducial mass & 0.04 & 3.7 \\
        Energy reconstruction & 0.02 & 1.8 \\
        Live time & 0.01 & 0.9\vspace{1mm} \\ \hline
        \vspace{-3.8mm} \rule{0pt}{1pt} & & \\
        \rule{0pt}{1pt}Total & 0.05 & 4.6 \\
        \br
        \end{tabular}
    \end{indented}
\end{table}
\normalsize

A first measurement of the relative capture fractions is performed by introducing an additional parameter for the KK fraction, allowing the two signal peaks to float separately. Relative ratios of modes comprising the KX peak were kept fixed. Bounds at $[0, 0.977]$ were imposed on this value, such that the 2.3\% proportion of $^{124}$Xe signal below the ROI is preserved, and the KX fraction can be inferred directly; the KK fraction was otherwise left unconstrained. The fitted KK and KX fractions are \qty{64.8(53)}{\%} and \qty{32.9(53)}{\%}, with an associated overall $\chi^{2}_{\lambda}/N_{\text{dof}} = 4843.42/4461 = 1.09$. At \qty{1.4}{\sigma} away from the values in table \ref{table:fractions}, these fit results are reasonably consistent with the assumed model. Future LZ runs will enable more precise measurements of these fractions, and may provide sensitivity towards the substructure of the KX peak.

\begin{figure}[!htb]
    \centering
    \includegraphics[width=\linewidth]{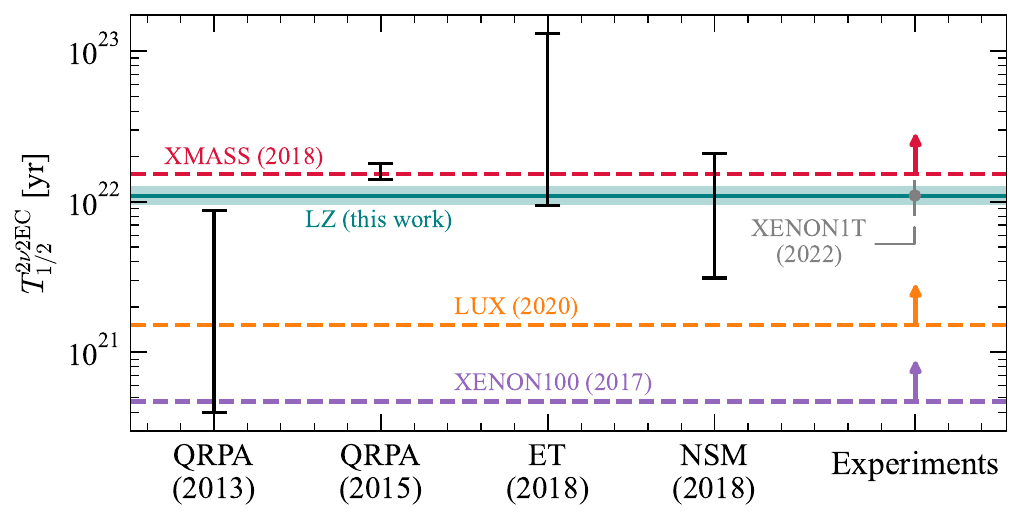}
    \caption{Comparison of the measured half-life with theoretical predictions represented by black bars and experimental lower limits indicated by dashed lines.}
    \label{fig:comparisons}
\end{figure}

The measured half-life is compared with recent predictions from various theoretical frameworks, as well as measurements and lower limits set by other experiments, in figure~\ref{fig:comparisons}. The result obtained here is compatible with effective theory (ET) and large-scale nuclear shell model (NSM) calculations \cite{ET_NSM}, and is in agreement with those from the quasiparticle random-phase approximation (QRPA) \cite{QRPA, QRPA_2015} approach at the \qty{2}{\sigma} level. Moreover, this measurement is also consistent with the 90\% confidence level lower limits set by XENON100 \cite{XENON100_Xe124} and LUX \cite{LUX_Xe124}, though a discrepancy is observed with respect to the limit produced by XMASS \cite{xmass_Xe124}.

% -------------------------------------------------------------

\section{Conclusion and outlook}

In this paper, we reported on the search for 2$\nu$2EC of $^{124}$Xe in the first science run of the LZ experiment. With an isotopic exposure of $\qty{1.39}{kg \times yr}$, we observed a half-life of $T_{1/2}^{\hspace{0.3mm}2\nu2\mathrm{EC}} = (1.09 \pm 0.14_{\text{stat}} \pm 0.05_{\text{sys}}) \times \qty{e22}{yr}$ at a significance of \qty{8.3}{\sigma}, demonstrating the versatility of LZ in exploring physics beyond its primary goal of searching for dark matter. We also report on the first measurement of the relative capture fractions for this decay, at \qty{64.8(53)}{\%} and \qty{32.9(53)}{\%} for the KK and summed (KL + KM + KN) modes respectively, which are relatively consistent with the adopted signal model \cite{XENON1T_DEC}. Additional data from upcoming LZ exposures will allow for these capture fractions to be measured with higher precision, and will ultimately enable searches for rarer variants of $^{124}$Xe decay, namely $2\nu\mathrm{EC}\beta^+$ and $2\nu2\beta^+$, along with their neutrinoless counterparts \cite{Wittweg}. Moreover, it may be of interest to also probe for enhancements to recombination in $^{124}$Xe decays in future work, as has been observed for single electron captures \cite{XELDA}.

% -------------------------------------------------------------

\section*{Acknowledgments}

The research supporting this work took place in part at the Sanford Underground Research Facility (SURF) in Lead, South Dakota. Funding for this work is supported by the U.S. Department of Energy, Office of Science, Office of High Energy Physics under Contract Numbers DE-AC02-05CH11231, DE-SC0020216, DE-SC0012704, DE-SC0010010, DE-AC02-07CH11359, DE-SC0015910, DE-SC0014223, DE-SC0010813, DE-SC0009999, DE-NA0003180, DE-SC0011702, DE-SC0010072, DE-SC0006605, DE-SC0008475, DE-SC0019193, DE-FG02-10ER46709, UW PRJ82AJ, DE-SC0013542, DE-AC02-76SF00515, DE-SC0018982, DE-SC0019066, DE-SC0015535, DE-SC0019319, DE-SC0024225, DE-SC0024114, DE-AC52-07NA27344, \& DE-SC0012447. This research was also supported by U.S. National Science Foundation (NSF); the UKRI’s Science \& Technology Facilities Council under award numbers ST/W000490/1, ST/W000482/1, ST/W000636/1, ST/W000466/1, ST/W000628/1, ST/W000555/1, ST/W000547/1, ST/W00058X/1, ST/X508263/1, ST/V506862/1, ST/X508561/1, ST/V507040/1 , ST/W507787/1, ST/R003181/1, ST/R003181/2,  ST/W507957/1, ST/X005984/1, ST/X006050/1; Portuguese Foundation for Science and Technology (FCT) under award numbers PTDC/FIS-PAR/2831/2020; the Institute for Basic Science, Korea (budget number IBS-R016-D1); the Swiss National Science Foundation (SNSF) under award number 10001549. This research was supported by the Australian Government through the Australian Research Council Centre of Excellence for Dark Matter Particle Physics under award number CE200100008. We acknowledge additional support from the UK Science \& Technology Facilities Council (STFC) for PhD studentships and the STFC Boulby Underground Laboratory in the U.K., the GridPP~\cite{faulkner2005gridpp,britton2009gridpp} and IRIS Collaborations, in particular at Imperial College London and additional support by the University College London (UCL) Cosmoparticle Initiative, and the University of Zurich. We acknowledge additional support from the Center for the Fundamental Physics of the Universe, Brown University. K.T. Lesko acknowledges the support of Brasenose College and Oxford University. The LZ Collaboration acknowledges the key contributions of Dr. Sidney Cahn, Yale University, in the production of calibration sources. This research used resources of the National Energy Research Scientific Computing Center, a DOE Office of Science User Facility supported by the Office of Science of the U.S. Department of Energy under Contract No. DE-AC02-05CH11231. We gratefully acknowledge support from GitLab through its GitLab for Education Program. The University of Edinburgh is a charitable body, registered in Scotland, with the registration number SC005336. The assistance of SURF and its personnel in providing physical access and general logistical and technical support is acknowledged. We acknowledge the South Dakota Governor's office, the South Dakota Community Foundation, the South Dakota State University Foundation, and the University of South Dakota Foundation for use of xenon. We also acknowledge the University of Alabama for providing xenon. For the purpose of open access, the authors have applied a Creative Commons Attribution (CC BY) license to any Author Accepted Manuscript version arising from this submission. Finally, we respectfully acknowledge that we are on the traditional land of Indigenous American peoples and honor their rich cultural heritage and enduring contributions. Their deep connection to this land and their resilience and wisdom continue to inspire and enrich our community. We commit to learning from and supporting their effort as original stewards of this land and to preserve their cultures and rights for a more inclusive and sustainable future.

% -------------------------------------------------------------

\section*{References}
\bibliography{references}

\providecommand{\newblock}{}
\begin{thebibliography}{10}
\expandafter\ifx\csname url\endcsname\relax
  \def\url#1{{\tt #1}}\fi
\expandafter\ifx\csname urlprefix\endcsname\relax\def\urlprefix{URL }\fi
\providecommand{\eprint}[2][]{\url{#2}}

\bibitem{benetti}
Benetti P {\em et~al.\/} 1993 {\em Nucl. Instrum. Methods Phys. Res. A\/} {\bf 332} 395--412

\bibitem{cline}
Cline D {\em et~al.\/} 2000 {\em Astropart. Phys.\/} {\bf 12} 373--377

\bibitem{majorana}
Majorana E 1937 {\em Nuovo Cim.\/} {\bf 14} 171--184

\bibitem{delloro_rev}
Dell’Oro S {\em et~al.\/} 2016 {\em Adv. High Energy Phys.\/} {\bf 2016} 2162659

\bibitem{dolinski_rev}
Dolinski M~J, Poon A~W and Rodejohann W 2019 {\em Annu. Rev. Nucl. Part. Sci.\/} {\bf 69} 219--251

\bibitem{agostini_rev}
Agostini M {\em et~al.\/} 2023 {\em {Rev. Mod. Phys}\/} {\bf 95} 025002

\bibitem{bernabeu}
Bernabeu J, De~Rujula A and Jarlskog C 1983 {\em {Nucl. Phys. B}\/} {\bf 223} 15--28

\bibitem{sujkowski_0v2EC}
Sujkowski Z and Wycech S 2004 {\em {Phys. Rev. C}\/} {\bf 70} 052501

\bibitem{vsimkovic}
{\v{S}}imkovic F, Krivoruchenko M and Faessler A 2011 {\em {Prog. Part. Nucl. Phys.}\/} {\bf 66} 446--451

\bibitem{Wittweg}
Wittweg C {\em et~al.\/} 2020 {\em {Eur. Phys. J. C}\/} {\bf 80} 1161

\bibitem{suhonen}
Suhonen J and Civitarese O 1998 {\em {Phys. Rep.}\/} {\bf 300} 123--214

\bibitem{HL_2v2EC}
Kotila J and Iachello F 2013 {\em {Phys. Rev. C}\/} {\bf 87} 024313

\bibitem{Ba130a}
Meshik A~P {\em et~al.\/} 2001 {\em {Phys. Rev. C}\/} {\bf 64} 035205

\bibitem{Ba130b}
Pujol M {\em et~al.\/} 2009 {\em {Geochim. Cosmochim. Acta}\/} {\bf 73} 6834--6846

\bibitem{baksan1}
Gavrilyuk Y~M {\em et~al.\/} 2013 {\em {Phys. Rev. C}\/} {\bf 87} 035501

\bibitem{baksan2}
Ratkevich S~S {\em et~al.\/} 2017 {\em {Phys. Rev. C}\/} {\bf 96} 065502

\bibitem{XENON1T_Xe124_KK}
Aprile E {\em et~al.\/} (XENON Collaboration) 2019 {\em {Nature}\/} {\bf 568} 532--535

\bibitem{XENON1T_DEC}
Aprile E {\em et~al.\/} (XENON Collaboration) 2022 {\em {Phys. Rev. C}\/} {\bf 106} 024328

\bibitem{LZ_NIM}
Akerib D {\em et~al.\/} (LZ Collaboration) 2020 {\em {Nucl. Instrum.}\/} {\bf 953} 163047 ISSN 0168-9002

\bibitem{LZ_TDR}
Mount B~J {\em et~al.\/} (LZ Collaboration) 2017 {LUX-ZEPLIN (LZ) Technical Design Report}

\bibitem{sanford}
Mei D~M {\em et~al.\/} 2010 {\em {Astropart. Phys}\/} {\bf 34} 33--39

\bibitem{muons}
Kudryavtsev V~A 2009 {\em {Comput. Phys. Commun.}\/} {\bf 180} 339--346

\bibitem{LZ_assays}
Akerib D~S {\em et~al.\/} (LZ Collaboration) 2020 {\em {Eur. Phys. J. C}\/} {\bf 80}

\bibitem{LZ_calibs}
Aalbers J {\em et~al.\/} (LZ Collaboration) 2024 {\em J. Instrum.\/} {\bf 19} P08027

\bibitem{LZ_SR1}
Aalbers J {\em et~al.\/} (LZ Collaboration) 2023 {\em {Phys. Rev. Lett.}\/} {\bf 131} 041002

\bibitem{W_val}
Szydagis M {\em et~al.\/} 2021 {\em {Instrum.}\/} {\bf 5} 13

\bibitem{DD_brown}
Verbus J~R {\em et~al.\/} 2017 {\em {Nucl. Instrum. Methods Phys. Res. A}\/} {\bf 851} 68--81

\bibitem{DD_LUX}
Akerib D~S {\em et~al.\/} (LUX Collaboration) 2016 {\em {arXiv:1608.05381}\/}

\bibitem{LZ_SR1BG}
Aalbers J {\em et~al.\/} (LZ Collaboration) 2022 {\em {Phys. Rev. D}\/} {\bf 108} 012010

\bibitem{Xe124_Qval}
Nesterenko D~A {\em et~al.\/} 2012 {\em {Phys. Rev. C}\/} {\bf 86} 044313

\bibitem{LUX_Xe124}
Akerib D~S {\em et~al.\/} (LUX Collaboration) 2020 {\em {J. Phys. G}\/} {\bf 47} 105105

\bibitem{Xe_IUPAC}
De~Laeter J~R {\em et~al.\/} 2003 {\em {Pure Appl. Chem.}\/} {\bf 75} 683--800

\bibitem{Xe_abundance}
Valkiers S {\em et~al.\/} 1998 {\em {Int. J. Mass Spectron.}\/} {\bf 173} 55--63

\bibitem{A133}
Khazov Y, Rodionov A and Kondev F 2011 {Nuclear Data Sheets for A = 133}

\bibitem{A127}
B{\'e} M~M {\em et~al.\/} 2016 {Table of radionuclides (Vol. 8 -- A = 41 to 198)}

\bibitem{A129}
Timar J, Elekes Z and Singh B 2014 {Nuclear Data Sheets for A = 129}

\bibitem{A125}
Katakura J 2011 {Nuclear Data Sheets for A = 125}

\bibitem{atomic_datasheet}
B{\'e} M~M {\em et~al.\/} 2011 {Table of radionuclides (Vol. 6 -- A = 22 to 242)}

\bibitem{Xray_trans1}
Deslattes R~D {\em et~al.\/} 2003 {\em {Rev. Mod. Phys.}\/} {\bf 75} 35--99

\bibitem{Xray_trans2}
Thompson A {\em et~al.\/} 2009 {X-ray data booklet}

\bibitem{LZ_sims}
Akerib D~S {\em et~al.\/} (LZ Collaboration) 2021 {\em {Astropart. Phys.}\/} {\bf 125} 102480

\bibitem{chisq}
Baker S and Cousins R~D 1984 {\em {Nucl. Instrum.}\/} {\bf 221} 437--442 ISSN 0167-5087

\bibitem{ET_NSM}
Coello~P\'erez E~A, Men\'endez J and Schwenk A 2019 {\em {Phys. Lett. B}\/} {\bf 797} 134885

\bibitem{QRPA}
Suhonen J 2013 {\em {J. Phys. G}\/} {\bf 40} 075102

\bibitem{QRPA_2015}
Pirinen P and Suhonen J 2015 {\em {Phys. Rev. C}\/} {\bf 91} 054309

\bibitem{XENON100_Xe124}
Aprile E {\em et~al.\/} (XENON Collaboration) 2017 {\em {Phys. Rev. C}\/} {\bf 95} 024605

\bibitem{xmass_Xe124}
Abe K {\em et~al.\/} (XMASS Collaboration) 2018 {\em {Prog. Theor. Exp. Phys.}\/}  053D03

\bibitem{XELDA}
Temples D~J {\em et~al.\/} 2021 {\em {Phys. Rev. D}\/} {\bf 104} 112001

\bibitem{faulkner2005gridpp}
Faulkner P~J~W {\em et~al.\/} 2005 {\em {J. Phys. G}\/} {\bf 32} N1

\bibitem{britton2009gridpp}
Britton D {\em et~al.\/} 2009 {\em {Philos. Trans. R. Soc. A}\/} {\bf 367} 2447--2457

\end{thebibliography}

\end{document}